%% file: ms_rev_v3.tex
\documentclass[referee]{aa}
\usepackage{graphicx,natbib}
\usepackage[latin1]{inputenc}
\usepackage[T1]{fontenc}
\bibpunct{(}{)}{;}{a}{}{,}
\usepackage{txfonts}

\include{def}
\usepackage{changebar}
\usepackage{bm}
\usepackage{units}
\usepackage{color}

\begin{document}

\title{Formation and composition of planets around very low mass stars}

\author{Y. Alibert \inst{1} \& W. Benz \inst{1}}
\offprints{Y. Alibert}
\institute{Physikalisches Institut  \& Center for Space and Habitability, Universitaet Bern, CH-3012 Bern, Switzerland, 
        \email{yann.alibert@space.unibe.ch,willy.benz@space.unibe.ch}
}

\abstract
{The recent detection of planets around very low mass stars raises the question of the formation, composition and potential habitability of these objects.}
{We use planetary system formation models to infer the properties, in particular their radius distribution and water content, of planets that may form  around stars ten times less massive than the Sun.}
{Our planetary system formation and composition models take into account the structure and evolution of the protoplanetary disk, the planetary mass growth
by accretion of solids and gas, as well as  planet-planet, planet-star and planet-disk interactions.}
{We show that planets can form at small orbital period in orbit about low mass stars. We show that the radius of the planets is peaked at about 
1 $\rearth$ and that they are, in general, volatile rich especially if proto-planetary discs orbiting this type of stars are long-lived. }
{Close-in planets  orbiting low-mass stars similar in terms of mass and radius to the ones recently detected can be formed within the framework of the core accretion paradigm as modeled here.
 The properties of protoplanetary disks, and their correlation with the stellar type, are key to understand their composition.}


\maketitle

\section{Introduction}
\label{sec:introduction}

The recent discovery of planets  orbiting low mass stars, as TRAPPIST-1 (Gillon et al. 2016) and Proxima Centauri b (Anglada-Escude et al. 2016),
as well as the non detection of similar planets by K2 (Demory et al. 2016) raises the question of the properties of planets forming around very low mass stars. 
Planets forming in such environment are specially interesting, for a number of reasons, {as recognised since a few years 
(e.g. Berta et al. 2012 for M stars, Belu et al. 2013, He et al. 2016 for brown dwarf) planets forming in such environment are specially interesting, for a number of reasons.}.
First, the detection and characterization of planets orbiting low mass stars is in 
general easier, as the Doppler effect and the transit depth (for transiting planets) are larger, for planets of similar mass. Second, as the equilibrium temperature scales 
with the stellar luminosity, planets in the so-called habitable zone are located closer to the star, and at shorter period. Finally, as we will see 
in this paper and a companion one (Alibert et al., in prep, hereafter paper II), the planet  formation process depends on the mass of the central star, either directly, or via 
the correlation between stellar type and protoplanetary disk properties. The composition of formed planets could therefore be substantially different around low-mass
stars than around solar type stars.

In this paper, we use our population synthesis models (Alibert et al. 2005, Fortier et al. 2013, Alibert et al. 2013, Benz et al. 2014, hereafter A13), adapted to the case of very low mass stars, 
to study the formation and the characteristics of planets in this environment. In this Letter, we focus on planets at short period that is orbiting within 0.1 au which
typically includes the habitable zone for these stars, and we concentrate on the bulk properties of planets (mass, radius, period, water content). A more detailed analysis on 
the formation process of these planets, as well as prediction on their potential detection by present and future surveys is the subject of paper II.
We recall  in Sect. \ref{model} the physics included in our models, 
 and present results of  computations, for planets that end their formation
inside 0.1 au of the central star (corresponding to a period of roughly 30 days) in Sect. \ref{formation}. The water fraction,
radius distribution and its their consequences on habitability are presented in  Sect. \ref{water}, and conclusions are
in Sect. \ref{conclusion}.

\section{Models}
\label{model}
\subsection{Formation}

Our planet formation models are based on the work of A13, adapted to the case of low mass stars. The models are described in detail in A13 and 
Fortier et al. (2013), and take into account different processes. The disk structure and evolution is computed in the framework of the $\alpha$ model by solving the diffusion
equation, the thermodynamical conditions in the disk midplane result from the
computation of the disk vertical structure. The nominal value of $\alpha$ is taken
to be  $2 \times 10^{-3}$ in this Letter. The effect of other values, in the  range  from $10^{-3}$ to
$10^{-2}$, is presented in paper II.
We consider different assumptions regarding the effect of irradiation by the central star: some models
are computed ignoring this effect, some are computed taking it into account. For irradiated models, we assume that the
flaring angle takes its equilibrium value (9/7, see Hueso and Guillot 2005).
The migration of the planet is computed using the formula presented
in Dittkrist et al. (2014) {which is similar to the one by Paardekkoper et al. (2010) approach, and allows outwards migration
in some circumstances.} 
{We do not take into account tidal interactions with the star, as they are only important for very close-in planets (see Bolmont et al., 2012). As 
a consequence, the starting time of our simulations has little effect (it would only modifies the irradiation from the star in irradiated
models - see below).}
{The planetary embryos grow as a result of planetesimal accretion. The planetesimal accretion rate depends on the planet
properties (for example more massive planets excite planetesimals, lowering the gravitational focussing and therefore the accretion rate,
see Fortier et al. 2013). Note that at the beginning of the simulation, the majority of the mass is in the planetesimals,
and not in the planetary embryos. The initial amount of planetesimal depends on the local gas surface density, and of the disk metallicity.
Finally, we do not take into account in this calculation shepherding of planetesimals (Tanaka \& Ida 1999), or planetesimal radial
drift (Thiabaud \& Alibert, in prep).}
{Planetary embryos can also accrete gas, the accretion rate depending on the internal structure of the gas envelope.
This internal structure} is computed by solving the standard planetary structure equations. This computation 
provides the evolution of the planetary mass and composition as a function of time.
Finally, the planetary radius is computed using the method presented in Alibert (2014, 2016), and depends on the planetary
core composition (in particular its fraction of volatiles) as well as on the planetary equilibrium temperature.

The models are computed assuming that the mass of the central star is equal to 0.1 $\msol$.
The mass of the central star affects different important processes:
1- the structure of the protoplanetary disk is modified for lower mass stars, as the viscous heating
(which depends on the keplerian frequency) and the irradiation depend on the mass, radius and effective
temperature of the central object, 2- the radius {of the planetary envelope during the formation phase} is equal (for not too small planets) to a fraction of the Hill radius, which depends
on the mass of the central body ({as a consequence, gas accretion is a function of the mass of the central star}),
and 3- the distribution of the disk mass is modified compared to the case of solar type stars (see below)

\subsection{Initial conditions}
\label{initial_conditions}

We consider a series of a few hundreds to thousands of identical stars, with mass equal to $0.1 \msun$, radius equal to $1.004 \rsol$ and
an effective temperature of 2935K. These latter parameters are important for models including irradiation
from the central star, and are taken from the evolution tracks of Baraffe et al. (2015), at an age of 1 Myr. 
Around each of these stars, a protoplanetary disk is assumed to exist (we do not model the formation of this
disk, nor the formation of initial planetary embryos). The disk profiles follows the one already used in A13. 
where the disk parameters are derived from the observations of Andrews et al. (2010). 

The distribution of disk masses around low mass stars is poorly known, but can be somewhat constrained from observations
that show that the accretion rate onto the star scales with the square of the stellar mass (e.g. Natta et al. 2004). Assuming that the viscosity
parameter does not depend on the mass of the star, Alibert et al. (2011) showed that this latter correlation is well
reproduced for a distribution of the disk masses $\mdisk$ following
$ \mdisk \propto M_{\rm star}^{1.2}$.
We note that this correlation is consistent with the one derived by Andrews et al. (2013), who quote a linear relation between
stellar and disk masses, {and with the recent results by Pascucci et al. (2016), who quote an exponent from 1.3 to 1.9}.
The distribution of disk masses we use in this Letter is therefore similar to the one used in A13, Thiabaud et al. (2014), but scaled
down by a factor $0.1^{1.2}$. Practically, the disk profile is the same, the normalization of the surface density at 5 au being reduced by this numerical
factor. In order to test the effect of this scaling, we have also computed a population assuming that the mass of the disk is two times larger, decreasing 
the viscosity by a factor 2, to keep the accretion rate unchanged. The other parameters are unchanged. 

The disk lifetime is assumed to be independent of mass of the central star, and its distribution is therefore similar
to the one by A13:  the cumulative distribution of disk lifetimes decays exponentially
with a characteristic time of 2.5 Myr. However, some recent observations point towards a dependance of the disk lifetime on  stellar mass.
For example, Ribas et al. (2015) compared the disk lifetime around stars larger than $2 \msol$ versus stars smaller
than $2 \msol$ and found that the former have a significantly shorter lifetime. It is not clear if this trend can be extrapolated
to the low mass stars we consider here, but, in order to test this possibility, we have also computed a model with a disk lifetime ten times longer
(a value that would follow from a disk lifetime inversely proportional to  stellar mass). For this model, we considered only one planet per disk.

We have considered that 10 planets grow in the same protoplanetary disk.
The initial masses of the planetary embryos are equal to the mass of the Moon, and the initial location of the planetary
embryo is drawn at random, the probability distribution of this initial location being uniform in log, between an inner radius
of 0.01 au and an outer radius of 5 au. Test have shown that planets initially located outside 5 au do not reach the
innermost 0.1 au of the protoplanetary disk and that they do not influence for growth and migration of innermost planets
(in the case of 10 planets forming in the same disk). The choice of 10 planetary embryos (and not more) is dictated essentially by computer
time constraints, as the time needed to run a model (of the order of a few days per planetary systems) scales as the square of the number 
of planetary embryos. We have however run some tests considering 20 planetary embryos, and we haven't found significant differences compared to 
the case of only 10 planetary embryos. This results is consistant with A13, where it was demonstrated that the properties of planets larger than 5 $\mearth$
(for a 1.0 $\msun$ star) do not depend on the number of embryos, provided it is larger enough (10). In the case of formation around
0.1 $\msun$ stars, the properties of planets larger than 0.5 $\mearth$ should not depend on the number of embryos, when it
is larger than 10. A more in-depth discussion of the effect of the number of embryo will be presented in paper II.

\section{Results}

\subsection{Orbital and bulk properties of formed planets}
\label{formation}

\begin{figure}
  \center
  \includegraphics[width=0.35\textheight]{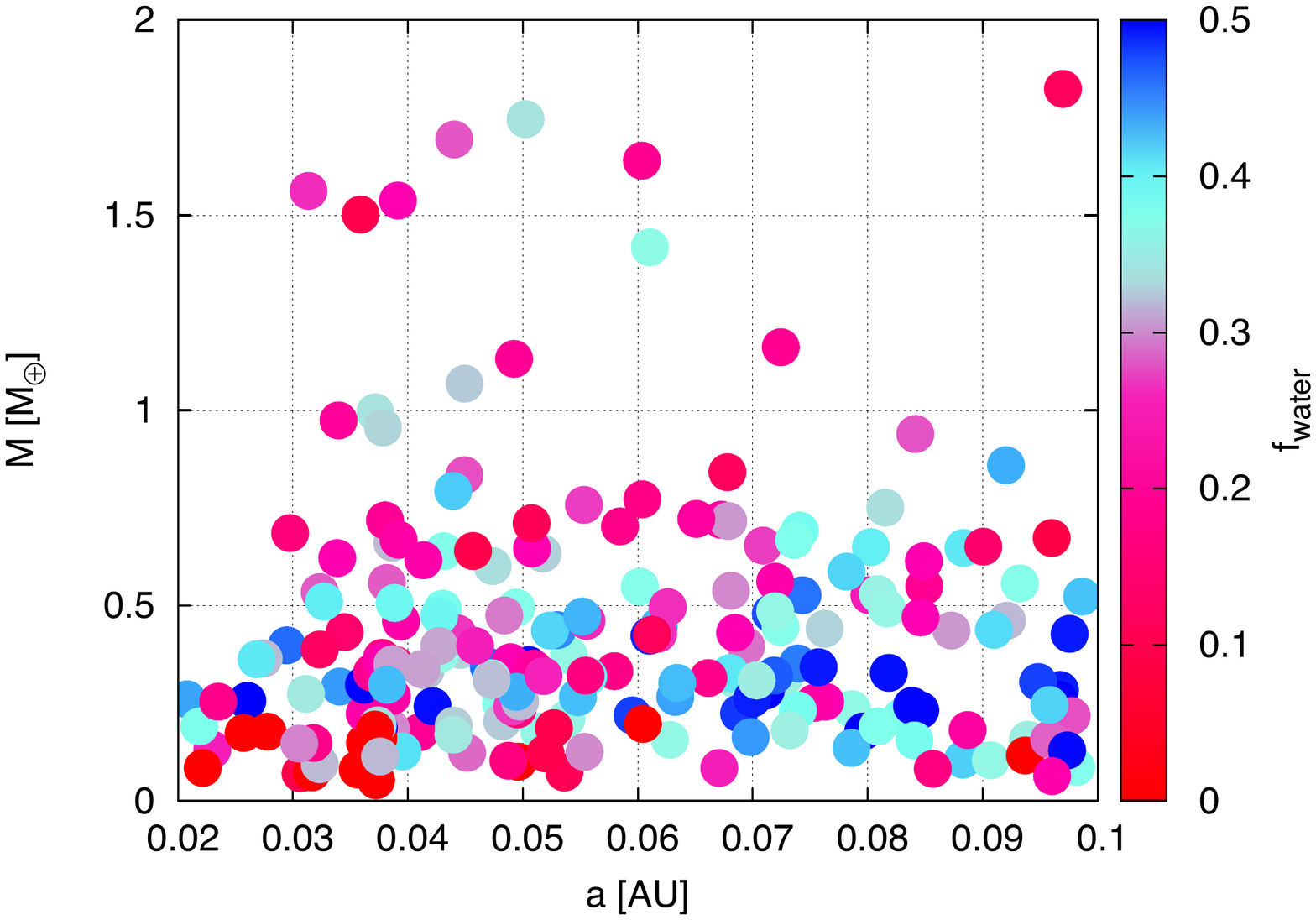}
  \caption{Mass \textit{versus} semi-major axis in our nominal model (disk mass $ \mdisk \propto M_{\rm star}^{1.2}$ and lifetime independent of stellar mass) .
  The color code gives the water mass fraction in the planet.}
  \label{am_water_10pla}
\end{figure}

Fig. \ref{am_water_10pla} shows the mass-semi-major axis distribution of planets located inside 0.1 au, in the case of
our nominal model.  The color code gives the fraction of water accreted in solid form by the planets ({see next section}). The distribution of masses and semi-major
axis is globally similar in the case where only one planet would form in a protoplanery disk. 

A detailed analysis of the populations, as well as some prediction on their observability, will be presented in paper II. Comparing
the simulations presented in this Letter, the population of planets is not strongly affected by the irradiation by the central star. This is
to be expected as the planets we consider form in the inner parts of the disk that is dominated by viscous heating. On the other hand, 
more important differences come from the massive disk population,  where more massive planets are produced, and from the long-lived 
disk population, where planets that end up inside 0.1 au, and do not collide  with the star, originate from larger distances. The first effect is 
simply due to the larger available mass, whereas the second effect is simply a result of the longer time available for migration: since the 
migration rate is similar in the nominal population and the one with longer disk lifetimes, planets that migrate within a given region of the 
disk (inside 0.1 au) must come from further out. As we will see below, this has important consequences on the water content of the planets.

We have also checked the rate of planetary system  forming in our simulations.  For this, we have considered all the systems that end with at least one planet more massive than
0.5 $\mearth$ closer than 0.1 au. In these systems,  $\sim 76 \%$ are single-planet systems, $16 \%$ are 2-planet systems, $6 \%$ are 3-planet systems,
and $2 \%$ are 4-planet systems (the statistic is however very low for this latter case). A more detailed study of the architecture of the systems, for the different
populations we have considered, will be presented in paper II.

\subsection{Water content, radius and consequence for habitability}
\label{water}

\begin{figure}
  \center
  \includegraphics[width=0.3\textheight,height=0.23\textheight]{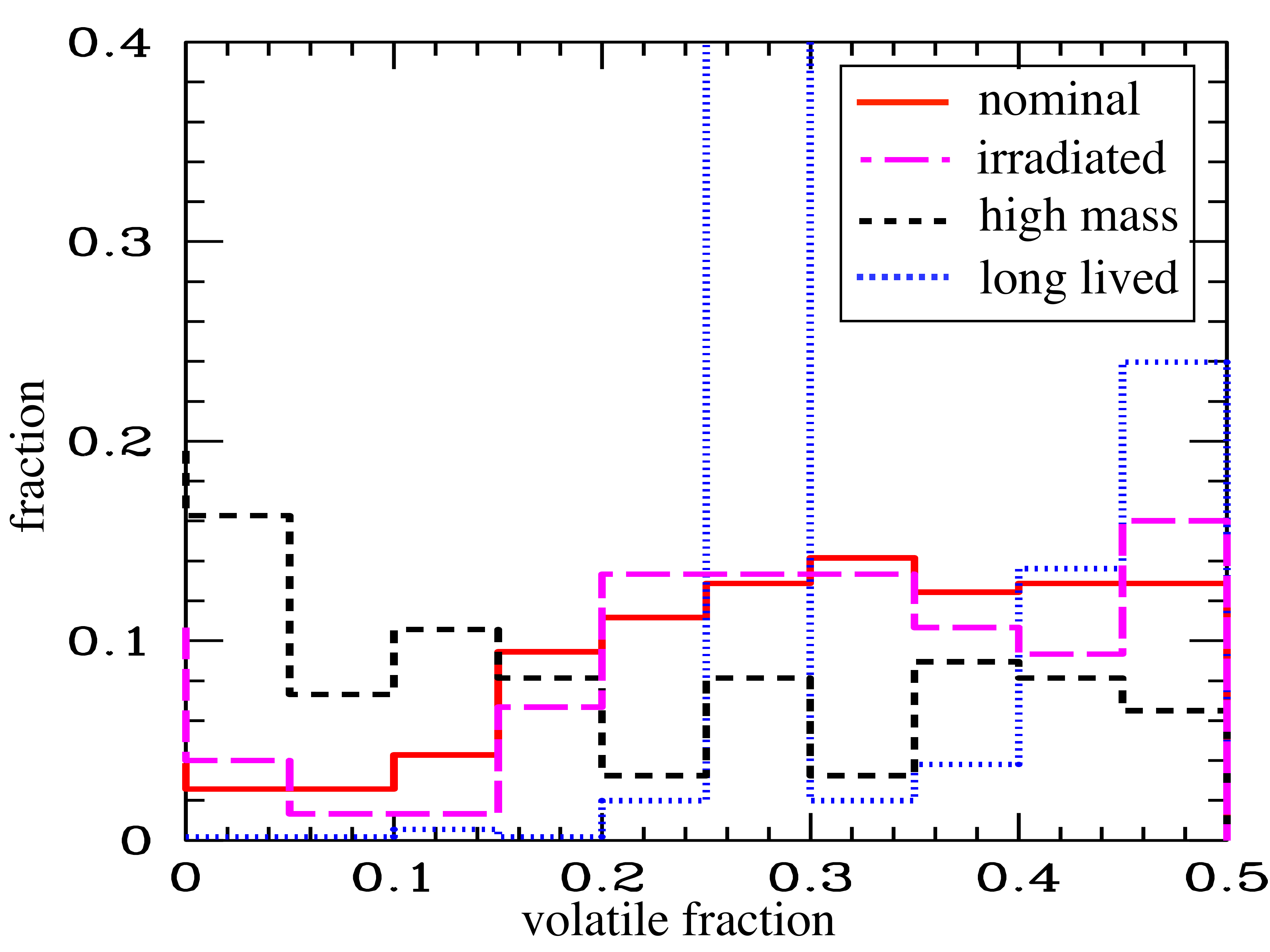}
  \caption{Distribution of volatile content, for the population indicated in the legend.}
  \label{histowater}
\end{figure}

{The amount of water depends upon the location at which the planet has accreted planetesimals whose composition
is determined by the thermal structure of the disk (in particular the location of the iceline) defined here as the innermost location of icy planetesimals}).
 {The location of the iceline depends on the thermal structure of disk, which in 
turn is a function of the disk mass, and ranges from  0.1 to 0.5 au in the majority of the cases. As radial drift and re-condensation of
water are negligible on large scales (Kornet et al., 2001), the location of the iceline does not evolve with time.} The fraction of volatiles accreted by the planets in the 
different populations is shown in Fig. \ref{histowater}. 
The distribution of water fraction is broad in all the cases, with a significant fraction of planets harboring more than 10 \% of water in all the cases. An interesting 
feature is that  planets forming in long lived disks all have a fraction of water larger than 10 \%. The reason for this large abundance is that planets have more time to 
migrate. The ones that end inside 0.1 au come therefore from beyond 0.5 au, where the temperature is lower than the condensation temperature of water, and they accrete 
during the beginning of their formation substantial amounts of icy planetesimals. In the case of massive disks {(disk masses being two times larger)}, the fraction of dry planets is much larger, and no planet have a 
water fraction larger than $\sim 35 \%$. This results from the fact that viscous heating scales with the mass of the disk. Massive disks are therefore much hotter,  the iceline 
is located at larger distance, and planets ending their formation within 0.1 au are  more water poor. 

In  the models we have considered, there is some correlation between the water fraction and the planetary mass:  planets that do not contain a lot of water are generally of
lower mass (below $\sim 1 \mearth$, although some can be as massive as $\sim 1.5 \mearth$), and planets totally devoid of water are all less
massive than $1 \mearth$, and even lower than $0.4 \mearth$ if one does not consider the massive disks. This can be easily understood as an effect of migration: more massive planets
migrate from further out and they collect some material outside the iceline (located between 0.1 and 1 au for the different models we have considered), and get enriched
in water.

We use the method by Alibert (2014, 2016) to compute the planetary radius. 
Interestingly enough, the majority of planets within 0.1 au have very small gaseous envelopes 
(of order 1 \% at maximum). We find this mass to be correlated with the total mass and the semi-major axis of the planet (see paper II for a more detailed discussion).
As this fraction is very small,
the planetary luminosity (or age) has a very small influence on the planetary radius. In order to save computer time, we have therefore
not computed the long-term thermodynamical evolution of the  planets, but we have assumed that the specific luminosity of these
planets follows the relation by Rogers et al. (2011). We have checked that increasing or decreasing the planetary luminosity by
one order of magnitude compared to the one given by Rogers et al. (2011) does not change the resulting radius.
The radius of the planetary core, on the other hand, does depend on the composition of the planet, in particular the fraction of volatiles,
as well as the Mg/Si and Mg/Fe ratios in the planet. 
Therefore in our computations we take into account the volatile content of each planet (shown in the previous section) while we assume the Mg/Si and Mg/Fe ratios to be solar. 
The volatile content in the planets results from the formation model as showed in the previous section.

\begin{figure}
  \center
  \includegraphics[width=0.3\textheight,height=0.21\textheight]{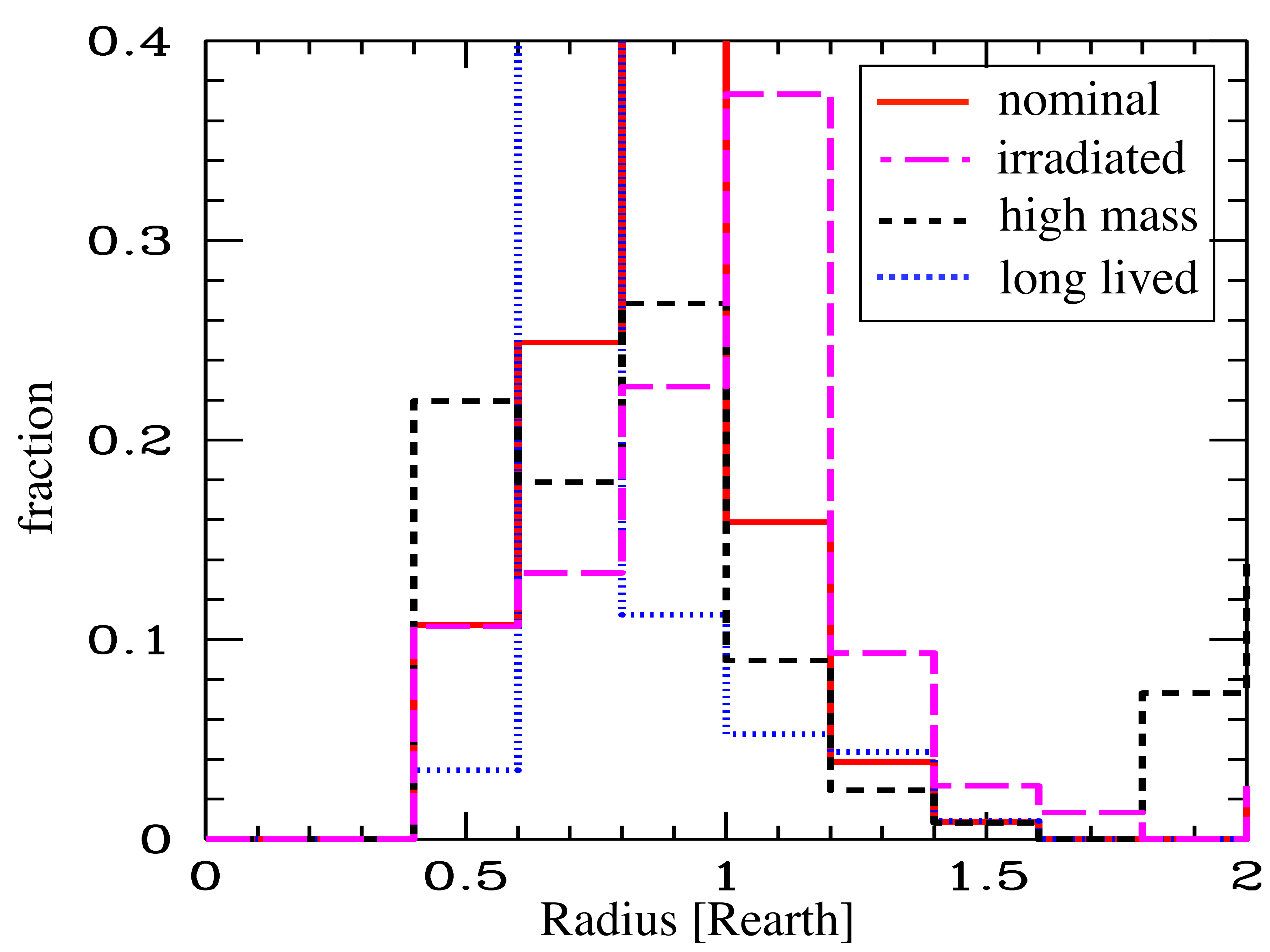}
  \caption{Histogram of radii, for the same cases as in Fig. \ref{histowater}.}
  \label{histoR}
\end{figure}

A clear feature of our models is a peak at $\sim 1 \rearth$ and a rapid cutoff at larger radii. This peak comes from the existence of a lower and upper limit
in the possible planetary mass.  First, at short distance to the star, the solid mass in the protoplanetary disk is small (in particular if the disk mass scales with 
the stellar mass). Planets of at least a fraction of an Earth mass must therefore have migrated from further out in the disk. As type I migration (the one that is 
important in this mass range) increases with planetary mass, only planets more massive than a fraction of an Earth mass can migrate from these
larger distances to within 0.1 au of the star. 
Second, even with migration, the total available mass to form planet is limited by the small mass of the disks.  It is 
therefore difficult to assemble a total mass larger than a couple of Earth masses, with a corresponding radius in the $1-1.5 \rearth$ range.
This explains the rapid decrease of the radius distribution at radii larger than $\sim 1.3 \rearth$. We note that this absence of larger planets is compatible with 
the non-detection by K2 of planets orbiting low mass M stars (see Demory et al. 2016). In the case of massive disks, a population of planets at larger radii appears
 (hints of this population are visible in Fig \ref{histoR}), and this population should have been detected by K2 (see Demory et al. 2016). 
The non detection of this population therefore tends to favor disk masses that follow the quasi-linear scaling between the disk and stellar mass.

The amount of water is important for habitability, as it is required for life as we know it, but, at the same time, 
too large amounts of water may be detrimental for habitability, at least for habitability as we know it on Earth, for at least two reasons. First, a too large water layer leads to the presence
of a high pressure ice layer at the bottom of the global ocean. This in turn prevents the existence of a {Carbonate-silicate cycle (Walker et al. 1981)} that could regulate the surface
temperature over long timescales (Alibert, 2014). In the case of low mass stars, which evolve on much longer timescales, this may not be a major problem,
as the stellar flux varies on timescales much longer than in the case of the Sun. In this situation, a process that stabilizes the surface temperature
may not be necessary. The second reason is connected to the fact that for planets with too much water an unstable CO$_2$ cycle destabilizes
the climate making habitability more challenging (Kitzmann et al. 2015)
 Again, this was demonstrated for solar-type stars and a similar process may or may not exist for low mass stars.

If, as it is the case for Solar type stars,  a large fraction of water does prevent habitability, the majority of planets formed in our models would not be habitable.
Indeed, the water mass fraction is in general larger than $\sim 10 \%$, and water evaporation has been claimed to be inefficient for planets in the 
habitable zone of low-mass stars (Ribas et al 2016).

 \section{Conclusion}
 \label{conclusion}
 
 Our models show that  planets quite similar in terms of mass and radius to the ones recently detected by 
(Gillon et al. 2016, Anglada-Escude et al. 2016) can be formed within the framework of the  core accretion paradigm as modelled here.
Moreover, the models predict a distribution of radius for these close-in planets relatively sharply peaked at about 1 $\rearth$. 

{Our results are consistant with the ones of Coleman et al. (2016), in particular their scenario (ii), which is similar to
the one we model here. On the other hand, other simulations concluded that planets forming around low mass stars should
be less massive and very dry (e.g. Lissauer et al. 2007, Raymond et al. 2007, Ciesla et al. 2015). The main reason for this difference
is that these later simulations consider the long term evolution of a swarm of massive planetesimals, after the gas disk has disappeared.
In this case, no migration of planets can occur, and planets located presently in the habitable zone accrete their mass inside the iceline,
resulting in a much lower water content. {The key role  of migration on the water content of planets around low mass stars
 that is demonstrated here is consistant with the results of Ogihara and Ida (2009).}}
    
   Finally, our models show that the properties of the disk and their potential correlation with the mass of the 
star are the most important parameters determining the characteristics, in particular the water content, of the emerging planet population. 
In this context, observational constraints  on mass and lifetime of discs in orbit of low-mass stars become of paramount importance. 
   
\acknowledgements

We thank M. Meyer and A. Dutrey for useful discussions. This work was supported in part by the European Research Council under grant 239605. This work has been carried out within the frame of the National Centre for Competence in Research PlanetS supported by the Swiss National Science Foundation. The authors acknowledge the financial support of the SNSF.

\end{document}

%% file: def.tex
\def\rearth{R_{\oplus}}

\def\f1{f_{\rm I}}
\def\msun{M_\odot}

\def\beq{\begin{equation}}
\def\eeq{\end{equation}}

\def\mdisk{M_{\rm disc}}

\def\mearth{M_\oplus}
\def\msun{M_\odot}

\def\mearth{M_\oplus}

\def\msol{M_\odot}

\def\simgr{\,\hbox{\hbox{$ > $}\kern -0.8em \lower 1.0ex\hbox{$\sim$}}\,}
\def\simle{\,\hbox{\hbox{$ < $}\kern -0.8em \lower 1.0ex\hbox{$\sim$}}\,}
\def\beq{\begin{equation}}
\def\eeq{\end{equation}}

\def\msol{M_\odot}

\def\simgr{\,\hbox{\hbox{$ > $}\kern -0.8em \lower 1.0ex\hbox{$\sim$}}\,}
\def\simle{\,\hbox{\hbox{$ < $}\kern -0.8em \lower 1.0ex\hbox{$\sim$}}\,}
\def\beq{\begin{equation}}
\def\eeq{\end{equation}}

\def\apj{ApJ}                 
\def\apjl{ApJ}                
\def\apjs{ApJS}               
\def\ao{Appl.Optics}          
\def\aap{A\&A}                
\def\mnras{MNRAS}             

\def\({\left(}
\def\){\right)}
\def\<{\left<}
\def\>{\right>}

\def\({\left(} 
\def\){\right)} 
\def\<{\left<} 
\def\>{\right>} 

\def\bc{\begin{changebar}}
\def\bce{\begin{center}}
\def\beq{\begin{equation}} 

\def\bi{\begin{itemize}}

\def\btab{\begin{tabular}{p{1.7cm}p{12cm}p{1.5cm}}}
\def\bt2{\begin{tabular}{p{1 cm}p{4.5cm}p{10cm}}}

\def\ec{\end{changebar}}
\def\ece{\end{center}}
\def\eeq{\end{equation}} 
\def\ei{\end{itemize}}

\def\etab{\end{tabular}\\}

\def\mH2{m_\mathrm{H_2}}

\def\dmax0{\rho_\mathrm{max}}
\def\dmaxS0{\Sigma_\mathrm{max}}

\def\msol{M_\odot} 

\def\rH2{r_\mathrm{H_2}}

\def\r0max{r_\mathrm{0max}}

\def\rsol{{\rm R}_\odot} 

\def\s0{\sigma_\mathrm{0}}

\def\xp0{x_{\rm{M0}}}

\def\z0max{z_\mathrm{0max}}

\def\apj{{\it ApJ}}                 
\def\apjl{ApJ}                
\def\apjs{ApJS}               
\def\ao{Appl.Optics}          
\def\aap{{\it A\&A}}                
 
\def\mnras{MNRAS}             